\documentclass{PoS}
  \usepackage{amssymb,amsmath,epsfig,bm,pifont}
\newcommand{\gev}[1]{\relax\ifmmode{\text{GeV}^{#1}}      
                     \else{GeV$^{#1}${ }}\fi}             
\def\Mev{\relax\ifmmode{\text{MeV}}\else{MeV{ }}\fi}      
\newcommand{\Ds}{\displaystyle}                           
  
\title{New extended Crewther-type relation}

\ShortTitle{New extended Crewther-type relation}

\author{\speaker{A.L.Kataev}\thanks{Second part  of the talk presented at 
at RADCOR2009- slightly improved version of preprint 
CERN-PH-TH/2009-203, by A.L. Kataev and S.V. Mikhailov, issued October 23, 
2009,  6 pages  
The first part, based on the work done in collaboratrion  with 
V.T.Kim (PNPI, Gatchina) will   be submitted  elswhere}\\
        Institute for Nucleaer Research of the 
Academy of Sciences of Russia, 117312, Moscow, Russia
        E-mail: \email{kataev@ms2.inr.ac.ru}}

\author{ S.V.Mikhailov 
       \\ Bogoliubov Laboratory of Theoretical Physics,
JINR, 141980, Dubna, Russia \\
       E-mail: \email{mikhs@theor.jinr.ru}}

\abstract{  We propose a conjecture about the detailed structure of the 
conformal symmetry breaking  term in the generalized  Crewther relation.
We conclude that this conjecture leads to new  relations between 
 the QCD expansion coefficients
of the Adler D-function and the polarized Bjorken sum rule B$_{jp}$.}

\FullConference{RADCOR 2009 - 9th International Symposium on Radiative Corrections (Applications of Quantum Field Theory to Phenomenology) ,\\
         October 25 - 30 2009\\
         Ascona, Switzerland}

\begin{document}

\section{Introduction}
The  concept of  conformal symmetry  is lying below definite modern 
theoretical and phenomenological investigations  in various massless quantum 
field models (for the most recent reviews see \cite{AGM99, BKM03}). 
Among them are 
quantum electrodynamics 
(QED) and quantum chromodynamics (QCD). Moreover,  
even before the formulation of QCD the study of  
axial-vector-vector (AVV) triangle amplitude in the conformal symmetry limit  
revealed   definite fundamental consequences of this  symmetry    
 \cite{CR}, \cite{ACGJ}. 
In particular 
the simple relation  between  basic characteristics of two main inclusive 
processes, 
namely  between flavour non-singlet parts of the 
Adler D-function $D_A^{NS}(a_s)$  and 
the non-singlet (NS) coefficient  function  $K(a_s)$, 
 which  enter into definitions of the Bjorken sum rule $S_{Bjp}$ of the 
polarized lepton-hadron 
 scattering and into the NS 
 contribution 
 of  the Gross-Llewellyn Smith sum rule $S_{GLS}^{NS}$ for the 
neutrino-nucleon scattering was found. 
 It can be written down in the following form  
 \begin{equation} \label{eq:CR}
D(a_s)\cdot K(a_s) = 1 
\end{equation}
 where $a_s=\alpha_s/(4\pi)$ and $\alpha_s$ is the QCD coupling.
 The entries in the LHS  
of this basic equation are defined as 
 \begin{eqnarray}
D^{NS}_{A}= \left(N_c \sum_f Q_f^2\right) \cdot D(a_s);~~~
S_{Bjp}=\left(\frac{1}{6}\frac{g_A}{g_V}\right)\cdot K (a_s)\,  ~~~S_{GLS}^{NS}=3\cdot K(a_s).
\end{eqnarray}
However, it is known that in the renormalized quantum field models   
conformal symmetry is broken by the procedure of renormalizations of the 
coupling constant  and the trace of  energy-momentum tensor.
In the first case the effect of conformal symmetry breaking is described by 
the renormalization-group 
$\beta$-function \cite{BSHbook}, while in the second case 
it leads to the  appearance of the conformal  anomaly
which is proportional to the factor  $\beta(a_s)/a_s$ 
\cite{EllisChanowitz,Nielsen,AdlerCollinsDunkan,CollinsDunkanJoglekar,
PMinkowski}.  Careful studies  of the structure of analytical results of NNLO 
perturbative QCD calculations of the  D-function 
\cite{GorKatL, SurgSam} and of the Bjorken polarized sum rule 
\cite{LarinVermaseren}, written down 
through Casimir color factors $C_F$, $C_A$ of the $SU_c(N)$ gauge group and the 
flavour factor $T_F N_F$, allowed to reveal that 
at the $a_s^3$-level the ``Crewther 
unity" is receiving additional conformal-symmetry breaking (CSB) contribution, 
\cite{BK93},    
\begin{equation} \label{eq:CI}
D\cdot K = 1 + \frac{\beta(a_s)}{a_s}~[\mbox{power series in}~a_s]
\end{equation}
which is proportional to the 
2-loop approximation of the  $\beta$-function. 
The application of the operator-product expansion method to the AVV 
diagram in the momentum 
space \cite{GabadadzeKataev} gave first indications, that  the factor 
$\beta(a_s)/a_s$ may be really factorized in   the RHS of Eq.(\ref{eq:CI}) 
in all orders of perturbation theory. 
This property was  proved later on in the coordinate space \cite{Cr97}, 
\cite{DMuller}. The subsequent  theoretical and 
 phenomenological studies of the 
 consequences of the discovered  in Ref. \cite{BK93} modification of the original Crewther relation 
were made in Refs. 
\cite{BrodskyGabadadzeKataevLu} and Ref.\cite{Rathsman}.

In this work we are proposing new  extension of the generalized 
Crewther relation of Refs.\cite{BK93}, \cite{Cr97}, specifying the structure of 
the power series in $a_s$ in the RHS of Eq.(\ref{eq:CI}). 
Using the method of the ``detailed $\beta$-expansion''  \cite{MS04}
we  are predicting some relations between $\alpha_s^4$  contributions 
to the $D$ and $K$-functions,  additional 
to those, which follow from the  original Crewther relation in QED  \cite{Kat08}.
The letter ones were  
already checked   by  the analytical calculations of   
the INR-Karlsruhe-SINP  collaboration \cite{BChK}.
We also emphasize the importance of knowledge of the   
complete analytical $a_s^4$  results of $D$ and $K$-functions    
for fixing some still unknown additional coefficients  
 in our new   relations, and at the first level  
in the  new extended Crewther-type relation. 
More detailed theoretical  explanations  are  
beyond the scope of this paper.

  \section{New extended  relation}
In this section we propose the following  new  ``conjectured" extended form of the
CSB term in the RHS  of Eq.(\ref{eq:CI}):
\begin{equation}
D(a_s)\cdot K(a_s) = 1 + \sum_n \left(\frac{\beta(a_s)}{a_s}\right)^{n}~{\cal P}_n(a_s)\, ,
\label{eq:newCI}
\end{equation}
where 
\begin{equation} 
{\cal P}_n(x):~~~~   
\Bigg\{
\begin{array}{l}
\Ds  {\cal P}_n(x)=x\ P_n\ C_F+ O(x^2) \\
 \mbox{polynomial}~{\cal P}_n(x)~~ \mbox{does not include the coefficients of~$\beta$-function} \nonumber \\
 \end{array}
\end{equation}
The coefficient functions $D(a_s)$ and $K(a_s)$ are taken in 
the $\rm{\overline{MS}}$-scheme at  $Q^2=\mu^2$, 
where $\mu$ is the renormalization  scale of the $\rm{\overline{MS}}$-scheme 
and 
\begin{eqnarray}
\mu^2 \frac{d}{d \mu^2} a_s=\beta(a_s)=-\left(b_0 a_s^2 + b_1 a_s^3+ b_2 a_s^4 +\ldots \right) 
\label{eq:beta}
\end{eqnarray}
At the level of order $O(a_s^3)$-corrections  Eq.(\ref{eq:newCI}) was   
obtained by requiring the independence of the  second term  of the power 
series in  the brackets of the RHS  of Eq.(\ref{eq:CI}) 
from  
the  first coefficient  $b_0$
of the  QCD $\beta$-function,  which both  contain  $T_FN_F$-terms given in  
Ref.\cite{BK93}. 
The concrete expressions for the polynomials ${\cal P}_n(a_s)$  
have the following form:  
\begin{eqnarray} 
\label{eq:P1}
{\cal P}_1(a_s)
&=& - a_s3C_F \left\{ \left(\frac{7}2-4\zeta_3 \right)+ a_s 
\left[ \left(\frac{47}{9}-\frac{16}{3}\zeta_3 \right)C_A-
\left(\frac{397}{18} + \frac{136}3\zeta_3 -80\zeta_5\right) C_F\right] \right. \nonumber \\ 
&& ~~~~~~~~~~~~~~~~~ +O(a_s^2) \bigg\} \\
{\cal P}_2(a_s) &=&a_s 3C_F \left(\frac{163}6-\frac{76}3\zeta_3\right)+O(a_s^2) \label{eq:P2} \\
 {\cal P}_3(a_s)&=& O(a_s) \label{eq:P3}
 \end{eqnarray} 
The results were  confirmed with the help ``detailed  $\beta$-function expansion'' 
formalism of  Ref.\cite{MS04}. 
The calculations of the order $O(a_s^4)$-corrections to $D$ and $K$-functions  
should give us 
the possibility to fix the  order  $O(a_s^3)$, $O(a_s^2)$ and 
$O(a_s)$- coefficients 
Eqs.(\ref{eq:P1}-\ref{eq:P3}) respectively. 
We expect that 
they will have  the following structure 
\begin{eqnarray}
\label{13}
 P_1^{(3)}(a_s)&=& a_s^3 C_F\bigg[C_F^2 {\rm {as_1}}+C_FC_A  {\rm{as_2}}+C_A^2 
{\rm{as_3}}\bigg] \\ \label{22}
 P_2^{(2)}(a_s)&=& a_s^2 C_F\bigg[C_F {\rm{as_4}}+ C_A {\rm{as_5}}\bigg] \\
\label{31}
 P_3^{(1)}(a_s)&=& a_s C_F ~{\rm{as_6}} 
\end{eqnarray} 
where ${\rm{as_i}}$ should contain   rational numbers  and 
the terms, proportional to $\zeta_3$, $\zeta_5$ 
and $\zeta_7$-functions, observed in the analytical calculations of order $O(a_s^4)$-correction to 
the Adler D-function \cite{BChK2}.

\section{Relations between  D and K coefficients  at $a_s^4$}    
Let us specify our expectations from the analytical $a_s^4$  calculations of 
$D$ and $K$ functions in a bit different form.
 We will define their perturbative as 
\begin{equation}
D=1 + \sum_n (a_s)^n~d_n, ~~~ K=1 + \sum_l (a_s)^l~k_l.
\label{eq: PT1}
\end{equation}
Following the analysis of \cite{MS04}, we consider an expansion of the
perturbative coefficients $d_n$ in a power series in $b_0, b_1,b_2,\ldots$,
as 
\begin{eqnarray}
d_2&=&b_0\,d_2[1]
  + d_2[0]\, ,\\
  d_3
&=&
  b_0^2\,d_3[2,0]
  + b_1\,d_3[0,1]
  + b_0\,d_3[1,0]
  + d_3[0,0]\, ,
\end{eqnarray}
as opposed to the standard expansion in a power series in $ N_F$,
$$d_3= N_F^2 \, d_3(2)+N_F^1\, d_3(1)+N_F^0\, d_3(0).$$
Here the first argument $n_0$ of the coefficients $d_n[n_0,n_1,\ldots]$
corresponds to the power of $b_0$, whereas the second one $n_1$
corresponds to the power of $b_1$, etc.
The coefficient $d_n[0,0,\ldots,0]$ represents ``genuine'' corrections
with powers $n_i=0$ for all the coefficients $b_i$.
If all the arguments of the coefficient $d_n[\ldots,m,0,\ldots,0]$
to the right of the index $m$ are equal to zero, then, for the sake of
a simplified notation, we will omit these arguments and write instead
$d_n[\ldots,m]$. 
Thus one can  write-down the analogous representation 
for the next coefficient $d_4$ in the following form: 
\begin{eqnarray}
 \label{eq:d_4}
  d_4
   &=& b_0^3\, d_4[3]
     + b_1\,b_0\,d_4[1,1]
     + b_2\, d_4[0,0,1]
     + b_0^2\,d_4[2]
     + b_1\,d_4[0,1]
     + b_0\,d_4[1]
     + d_4[0]\,. 
\end{eqnarray}
The same ordering in  the $\beta$-function elements applies for all
higher coefficients $d_n$ and for $k_l$ as well.
 The standard naive non-abelianization (NNA) approximation  estimates $d_n$
from the first term in the equations above, namely,
$d_n \simeq b_0^{n-1}d_n[n-1]$ (see Refs.\cite{LovettTurner:1995ti, Broadhurst:2002bi}).  Note that  the considerations 
of Ref.\cite{LovettTurner:1995ti}  are taking into account the 
terms in the coefficients  $d_n$, which are proportional to  $b_0^{i}$ 
with $1\leq i \leq (n-1)$, but the concrete expansions  
in higher order coefficients 
of $\beta$-function with  
$b_i (i\geq 1)$ are not included in this renormalon-inspired approach.

The conformal limit in (\ref{eq:newCI}) 
provides the relation between unknown yet elements  of 5-loop terms  $d_4, k_4$, 
and already known parts  of the 4-loop results.
Indeed, 
Eq.(\ref{eq:CR}) is satisfied at $\beta=0$, when all the coefficients have genuine content only,
$d_n~(k_n) = d_n[0]~(k_n[0])$,
that provides the evident relation between these elements at any loops
\begin{equation}
k_n[0] + d_n[0] + \sum_{l=1}^{n-1} d_l[0] k_{n-l}[0]=0\ . 
\label{eq:CI-PT0}
\end{equation}
In particular, 
this gives a relation of the results of  
5-loop calculations  $k_4[0], d_4[0]$ in the LHS of Eq.(\ref{eq:CI-PT0})
\begin{eqnarray} 
\label{eq:k4-d4}
&&k_4[0] + d_4[0]= 2d_1 d_3[0]-3d_1^2d_2[0]+(d_2[0])^2+d_1^4,
\end{eqnarray}
and the expressions  for $d_i[0]$ up to 4-loop level in its  RHS.
The explicit form of  $C_F$ and $C_A$-
dependence of the coefficients $d_2[0]$, $d_3[0]$, which enter into 
Eq.(\ref{eq:k4-d4}), is derived in Ref. \cite{MS04}.
Note, that the terms $d_4[0]$ and $k_4[0]$ should contain the dependence on 
two Casimir operators $C_F$ and $C_A$.  
In view of this the direct check of Eq.(\ref{eq:k4-d4}) is stronger than 
already performed in Ref.\cite{BChK} confirmation of the study of the 
 validity of its    
the projection onto  the maximum power of $C_F$, i.e. $C_F^4$,  
suggested in \cite{Kat08} as the  check of  the containing $\zeta_3$-term 
results for the QED part of  $d_4$-term, available from  publication 
in Ref.\cite{BChK3} 

Using  the conjecture of Eq.(\ref{eq:newCI}) 
one  may predict  others relations between the 
of $d_4~(d_n)$ and $k_4~(k_n)$ elements.  
For example,
in virtue of the $\beta$ factorization, 
one can put for the coefficient in front of $a_s^2 b_0, a_s^3 b_1, a_s^4 b_2, \ldots$,
(for the term $(\beta(a_s)/a_s)^1$ in (\ref{eq:newCI})) the chain of equations
\begin{eqnarray} \label{eq:k_n1-d_n1}
- P_1=k_2[1] + d_2[1] &=& k_3[0,1] + d_3[0,1]= k_4[0,0,1] + d_4[0,0,1]= \ldots  \nonumber \\
&=& k_n[0,0,\ldots, 1] + d_n[0,0,\ldots, 1]= 3C_F \left(\frac{7}2-4\zeta_3 \right)\, ,
 \end{eqnarray}
that fixes the  term $P_1$ in the  ${\cal P}_1$. 
The  term at $a_s^2$  in Eq.(\ref{eq:P1}) 
generates another   relation
at $a_s^4$ level
\begin{eqnarray} \label{k_401-d_401}
k_4[0,1] + d_4[0,1]= k_3[1] + d_3[1]+ d_1\left(k_{2}[1]-k_3[0,1] + d_3[0,1] -d_{2}[1]\right),
 \end{eqnarray}
where the RHS can be fixed after  applying the method of \cite{MS04} to 
the coefficients of the $S_{Bjp}$ as well.
However,  to use  this equation 
 in practise  we should know  the analytical  5-loop results for the   
$D^{NS}_{A}$ and $S_{Bjp}$ (for later presentation see  \cite{BChK4}).  

\acknowledgments
This work was started  at  Dubna in August of 2009, 
when both of us were participating at Bogolyubov Conference 
on Quantum Field Theory and Elementary Particle Physics. 
We are grateful to K.G. Chetyrkin for discussions during this Conference. 
We also wish to thank V.M.Braun and D.J. Broadhurst for useful communications 
prior and after this event. 
One of us (ALK) is grateful to his colleagues for creating 
rather stimulating atmosphere during his stay at Th Unit of  CERN up to  
24 October,   2009 and to the Members of Organizing Committee of 
RADCOR-2009 for the invitation,  hospitality and financial support
during his stay at this important Symposium.   
The work  of both of us was  supported in  part by   
the RFBR  grant  No.\ 08-01-00686, while 
the work of ALK in part  by the grant of President of 
RF NS-1616.2008.2 as well.

\section{Note added}
After our work and the one of Ref.\cite{BChK4} were subsequently presented 
and the validity 
of the generalized Crewther relation, written down in the form of 
Eq.(\ref{eq:CI}) was demonstrated  at the $\alpha_s^4$-level  
\cite{BChK4}, we checked that proposed by us  new    
representation for the conformal symmetry breaking term,   
 written down in the form of Eq.(\ref{eq:newCI}),  
is valid at the $\alpha_s^4$-level.
The explicit analytical 
expressions of the coefficients in Eqs.(\ref{13}-\ref{31}) were  
fixed directly  from the  analytical  
$\alpha_s^4$-expression for Eq.(\ref{eq:CI}), which were  
taken from the results of Ref.\cite{BChK4} only. 
It should be noted that after the talk of  Ref.\cite{BChK4} was reported 
we learned that these results were  presented  before  
RADCOR2009  as well  in the talk of Ref.\cite{BChK5}.
Unfortunately, we were unable to check our new guesses prior 
RADCOR2009 Workshop in view of the delay of the  kind information by 
Prof.K.G.Chetyrkin about the previous report. 
Thus, considering 
this fact as the ``closed-box'' check,       
we are  grateful to him for the real interest in our work.

\end{document}